\documentclass[journal]{IEEEtran}

\usepackage{ulem}
\usepackage{xcolor,soul,framed} 
\usepackage{amssymb,stmaryrd}   
\colorlet{shadecolor}{yellow}
\usepackage[pdftex]{graphicx}
\graphicspath{{../pdf/}{../jpeg/}}
\DeclareGraphicsExtensions{.pdf,.jpeg,.png}
\usepackage{bm}
\usepackage[cmex10]{amsmath}
\usepackage{array}
\usepackage{mdwmath}
\usepackage{mdwtab}
\usepackage{eqparbox}
\usepackage{url}
\usepackage{hyperref}
\usepackage{cite}
\usepackage{amsmath,amssymb,amsfonts}
\usepackage{algorithmic}
\usepackage{graphicx}
\usepackage{textcomp}
\usepackage{xcolor}
\usepackage{multirow} 
 \usepackage{booktabs}
 \usepackage{graphicx} 
 \usepackage{graphicx, caption, sidecap}
\usepackage{amssymb}
\def\BibTeX{{\rm B\kern-.05em{\sc i\kern-.025em b}\kern-.08em
    T\kern-.1667em\lower.7ex\hbox{E}\kern-.125emX}}
\usepackage{array, makecell} %
\usepackage{stmaryrd}
\usepackage{tikz}    
\usepackage{enumerate}
\usepackage{tikz}
\usepackage{amsmath}
\usetikzlibrary{positioning, calc, shapes.geometric}

\begin{document}
\bstctlcite{IEEEexample:BSTcontrol}
    \title{Goal-Oriented Source Coding using LDPC Codes for Compressed-Domain Image Classification}
  \author{Ahcen~Aliouat,~\IEEEmembership{Member,~IEEE,}
      and Elsa~Dupraz,~\IEEEmembership{Senior Member,~IEEE,}

  \thanks{This paper is an extended version of the paper ``Learning on JPEG-LDPC Compressed Images: Classifying with Syndromes" presented at the $32^{\text{nd}}$ edition of the EURASIP EUSIPCO 2024 conference. Here, we provide a deeper understanding of the reasons why the GRU+LDPC approach works well, and investigate the effect of various code and model parameters on the classification performance.  } \thanks{This work has received French government support granted to the Cominlabs excellence laboratory and managed by the National Research Agency in the `Investing for the Future program under reference ANR-10-LABX-07-01.}
  \thanks{Ahcen Aliouat and Elsa Dupraz are with MEE Departement, IMT Atlantique, CNRS UMR 6285, Lab-STICC, Brest, France (e-mail: ahcen.aliouet@imt-atlantique.fr, elsa.dupraz@imt-atlantique.fr).}
  %
}

\markboth{Submitted to IEEE TRANSACTIONS ON COMMUNICATIONS,  January~2025
}{Roberg \MakeLowercase{\textit{et al.}}: Goal-oriented bitplanes source-coding using LDPC for compressed-domain image classification}

\maketitle


\begin{abstract} 
In the emerging field of goal-oriented communications, the focus has shifted from reconstructing data to directly performing specific learning tasks, such as classification, segmentation, or pattern recognition, on the received coded data. In the commonly studied scenario of classification from compressed images, a key objective is to enable learning directly on entropy-coded data, thereby bypassing the computationally intensive step of data reconstruction.
Conventional entropy-coding methods, such as Huffman and Arithmetic coding, are effective for compression but disrupt the data structure, making them less suitable for direct learning without decoding. This paper investigates the use of low-density parity-check (LDPC) codes—originally designed for channel coding—as an alternative entropy-coding approach. It is hypothesized that the structured nature of LDPC codes can be leveraged more effectively by deep learning models for tasks like classification.
At the receiver side, gated recurrent unit (GRU) models are trained to perform image classification directly on LDPC-coded data. Experiments on datasets like MNIST, Fashion-MNIST, and CIFAR show that LDPC codes outperform Huffman and Arithmetic coding in classification tasks, while requiring significantly smaller learning models. Furthermore, the paper analyzes why LDPC codes preserve data structure more effectively than traditional entropy-coding techniques and explores the impact of key code parameters on classification performance. These results suggest that LDPC-based entropy coding offers an optimal balance between learning efficiency and model complexity, eliminating the need for prior decoding.
\end{abstract}


%
\IEEEpeerreviewmaketitle


\section{Introduction}
\IEEEPARstart{T}{he} paradigm of goal-oriented and semantic communications promises to be a key technology in sixth-generation (6G) telecommunication networks~\cite{strinati20216g,li2024toward}. 
Unlike in traditional communication schemes focused on data reconstruction, this paradigm considers the application of specific tasks over transmitted data~\cite{pappas2021goal,stavrou2023role,agheli2024semantic,talli2024effective,wang2024adaptive}. 
%
%
%
Within this framework, learning over compressed data has emerged as an important area of research driven by the increasing amount of visual data, such as images and videos, that must be stored, processed, and transmitted efficiently~\cite{kang2022task,shao2021learning}. In the field of data compression, this emerging paradigm is now referred to as Image and Video Coding for Machines (VCM) \cite{duan2020video}. VCM aims to develop source coding methods that support machine learning tasks such as classification or segmentation over compressed data. VCM is addressed in practice in the JPEG-AI initiative\cite{ascenso2023jpeg}, which aims to introduce compressed domain representations optimized for learning tasks. A key desired feature of VCM is to enable the application of learning tasks directly on compressed data without any prior decoding. 


Recent advances in learning over compressed data can be grouped into two main categories. The first one relies on end-to-end learning models that integrate deep neural networks into both the encoder and the decoder~\cite{balle2016end,balle2018variational,minnen2018joint,lee2023entropy,jeon2023context,jin2023learned}. With these models, often built using autoencoders, the receiver is trained to perform the task directly on a learned latent space~\cite{torfason2018towards,zhang2023machine}.
For instance,~\cite{cheng2018deep} introduces a model that classifies compressed images without prior decoding, while~\cite{enttselanomaly,enttsel2024enhancing} employ entropy-regularized autoencoders for tasks such as anomaly detection. However, these methods are not compatible with widely adopted image compression standards such as JPEG. In addition, they are computationally heavy, and they do not generalize well over unseen datasets~\cite{10337636}.

%


\begin{figure}[htbp]
\includegraphics[width=\linewidth]{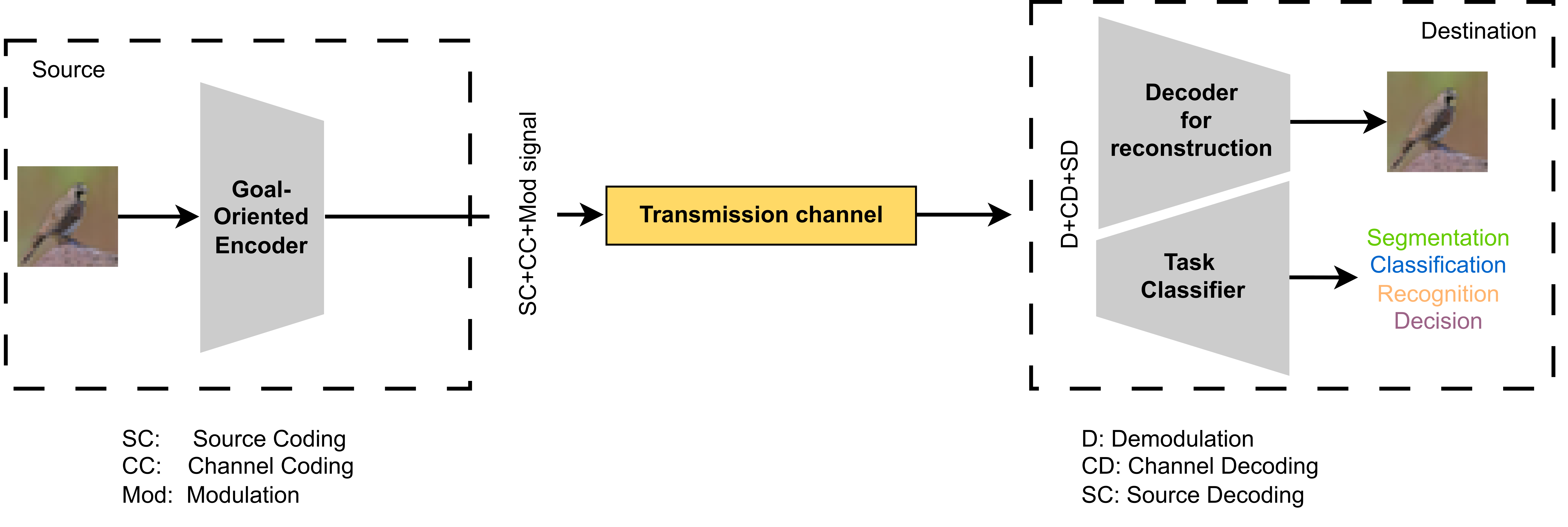}
\caption{\textbf{Goal-Oriented Communication}: this setup allows for both learning and reconstruction of compressed data.}
\label{GoC_setup}
\end{figure}

The second category considers applying the learning task over conventional compression standards. Notable examples include training convolutional neural networks (CNNs) directly on block-wise JPEG discrete cosine transform (DCT) coefficients~\cite{gueguen2018faster}. Learning accuracy can be further improved by selecting specific DCT coefficients~\cite{abdellatef2024reduced}, or employing vision transformers with feature patch embeddings~\cite{ji2024compressed}. However, these approaches often require partial decoding, such as reconstructing DCT coefficients, which adds complexity and latency. An important step toward bypassing partial decoding is introduced in~\cite{fu2016using,remy1,piau2023predicting}. Therein, established CNN-based models for image classification, such as ResNet and VGG, are trained directly on images compressed using standard Arithmetic and Huffman entropy coding techniques. Despite eliminating the need for decoding, these methods suffer from low classification accuracy due to the disruptive nature of conventional entropy coders on image structure, notably compromising spatial closeness between pixels.

In the case of video coding, studies such as~\cite{chamain2021end,tan2020real,duan2023unified} have explored machine vision tasks on compressed videos. However, these approaches face similar limitations, including incompatibility with conventional video coding standards (\emph{e.g.}, MPEG, HEVC), reliance on large neural network models, or the need for partial decoding.

In this paper, we propose a novel approach to image classification over compressed data that eliminates the need for any form of decoding. Building upon our earlier work~\cite{aliouat2024learning}, we explore alternative entropy-coding techniques, specifically Low-Density Parity Check (LDPC) codes, a family of channel codes which have also demonstrated strong performance in noiseless entropy coding~\cite{caire2004noiseless}, distributed source coding~\cite{liveris2002compression}, and 360-degree image compression~\cite{9319718}. Our rationale for examining this unconventional entropy-coding approach stems from the key idea that the intrinsic structure of LDPC codes may be used more effectively when training models for image classification. 
%
%
We consider standard JPEG compression, with its typical steps including DCT and quantization, while substituting the final entropic coder with LDPC codes applied over bitplanes. The latter step generates distinct codewords, termed syndromes, for each bitplane. At the receiver, we employ Recurrent Neural Networks (RNNs)~\cite{salehinejad2017recent}, specifically the Gated Recurrent Unit (GRU) model, to classify images based on their LDPC-coded bitplane syndromes. 
%
%
Our results demonstrate that this method achieves superior classification accuracy compared to existing approaches using Huffman or Arithmetic coding~\cite{remy1,piau2023predicting}. Furthermore, our methodology exhibits robust performance across various LDPC code parameters such as the code length, the rate, or the degree distribution. Additionally, our GRU-based model requires significantly fewer parameters than CNN-based models like VGG or ResNet, which shows its computational efficiency. 
Therefore, the proposed approach has the potential to streamline the learning pipeline by eliminating the need for decoding. It also offers a new perspective on the synergy between compression techniques and deep learning models.


The main contributions of this paper can be summarized as follows:
\begin{enumerate}
\item We investigate two setups for image classification over compressed images. In the first setup, LDPC codes are applied directly on the bitplanes of the images. In the second setup, we integrate LDPC coding into the standard JPEG pipeline by retaining the DCT and quantization steps while replacing the traditional entropy encoder with an LDPC encoder. In both setups, we employ a GRU model trained for classification directly over the resulting LDPC-coded data,  eliminating the need for prior decoding. 
\item  We demonstrate that GRU models trained on LDPC-coded data achieve up to $15\%$ higher classification accuracy compared to existing methods~\cite{remy1,piau2023predicting} that rely on Huffman or Arithmetic coding combined with CNN models such as VGG and ResNet. Furthermore, the proposed GRU-based approach significantly reduces model complexity by requiring far fewer parameters than the CNN-based methods.
\item  To better understand the effectiveness of our approach, we analyze the disorder introduced by LDPC coding using the notion of approximate entropy~\cite{piau2023predicting} as well as a t-SNR analysis. Our results reveal that LDPC coding preserves structural patterns in the data more effectively than conventional entropy coding techniques. Notably, LDPC-coded data exhibit consistent patterns across images of the same class, enhancing their predictability for learning models.  
\item We investigate the impact of various LDPC code parameters, such as code length, code rate, and degree distribution, as well as compression parameters including the number of bitplanes and JPEG quality factors. This analysis allows us to investigate the trade-offs between coding rate, classification accuracy, and model complexity, providing insights into the limits and scalability of the proposed approach.  
\end{enumerate}

The remainder of this paper is organized as follows. Section~\ref{sec:ldpc_codes} describes entropic coding using LDPC codes. Section~\ref{preliminary} introduces the proposed method for learning over compressed images, presenting both the proposed coding schemes and the classification model. Section~\ref{sec:data_exp}  describes the experimental setup, and Section~\ref{sec:results_disc} provides numerical results.



\section{Entropy coding with LDPC codes}\label{sec:ldpc_codes}



In this section, we describe how LDPC codes can be used as a source entropy coding technique applied over the bitplanes of an image. 

\subsection{Bitplane decomposition} 
\label{bp_decomp}


We first describe image bitplane decomposition, a prerequisite step before applying binary LDPC encoding. Note that bitplane decomposition was also considered in~\cite{piau2023predicting} when applying Huffman and arithmetic coding. 

Given an image \( \mathbf{I} \) of dimensions \( P \times Q \), where each pixel value \( I_{p,q} \), $p \in \llbracket 1,P\rrbracket$, $q\in  \llbracket 1,Q\rrbracket$, is a $K$-bit integer, the binary form of pixel $I_{p,q}$ is expressed as
\begin{equation}
    I_{p,q} = b_{p,q}^{(K-1)} b_{p,q}^{(6)} \cdots  b_{p,q}^{(0)},
\end{equation}
where $b_{p,q}^{(K)}$ is the most significant bit (MSB), and $b_{p,q}^{(0)}$ is the least significant bit (LSB). 
The expression of the $k$-th bit $b_{p,q}^{(k)}$, $k \in \llbracket 0,K-1 \rrbracket$, of the pixel $I_{p,q}$ is given by:
\begin{equation}
b^{(k)}_{p,q} = \left\lfloor \frac{I_{p,q}}{2^k} \right\rfloor \mod 2,
\label{eq_bp}
\end{equation}
where $\lfloor \cdot \rfloor$ denotes the floor function. 

Next, we use \( \mathbf{x_k} \) to denote the $k$-th bitplane of the image. 
For each $k \in \llbracket 0,K-1\rrbracket$, the bitplane \( \mathbf{x_k} \) is of length $N = P\times Q$ and it is constructed by extracting the $N$ $k$-th bits $b_{p,q}^{(k)}$ from the binary representations of the $N$ pixels $I_{p,q}$. 
 Therefore, each bitplane can be seen as a vectorized image in binary form, 
and the bitplane $\mathbf{x}_{K-1}$ constructed from the MSBs encompasses most of the image's information. 

\begin{figure*}[htbp]
\includegraphics[width=\linewidth]{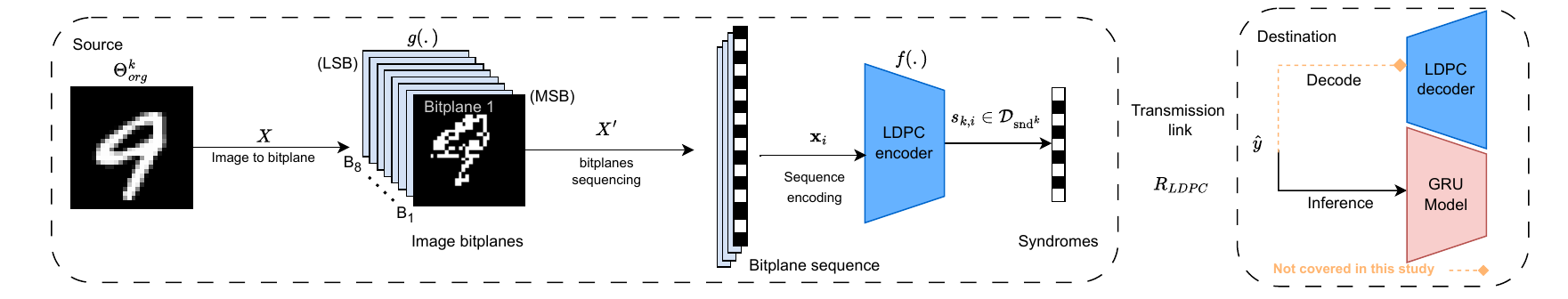}
\caption{\textbf{First setup}: Syndromes obtained from the LDPC-coded bitplanes are fed as input of a GRU model for classification}
\label{setup1}
\end{figure*}
\subsection{Source coding using LDPC codes}
We now describe the use of LDPC codes for entropy coding of the bitplanes, following an approach initially introduced in~\cite{caire2004noiseless} in the context of lossless source coding. 
First, we transform the $N$ image pixels into $K$ bitplanes  $\mathbf{x}_k$ as described in Section~\ref{bp_decomp}. %
We then consider binary LDPC codes, given that previous studies have shown that this does not significantly impact the coding rate compared to using non-binary LDPC codes directly over the pixels of the image~\cite{9319718}.

Let $H$ be the binary parity check matrix of size $M \times N$, where
$M < N$, of a binary LDPC code~\cite{liveris2002compression,ye2019optimized}. Assuming that $H$ is full rank,
the source coding rate is defined as $R = M/N$. 
In our scheme, each bit plane $\mathbf{x}_k$, $k \in \llbracket 0,K-1\rrbracket$, has the same length $N$ and it is compressed by an LDPC code using the following formula~\cite{liveris2002compression}:
\begin{equation}\label{eq:ldpc_enc}
    \mathbf{s}_k = H\mathbf{x}_k \mod 2,
\end{equation}
where $\mathbf{s}_k$ is a binary sequence of length $M$ called the syndrome. In our scheme, the $K$ syndromes form the codewords sent to the decoder. On the receiver side, these syndromes can be used either for data reconstruction as in~\cite{caire2004noiseless,liveris2002compression,9319718}, or as input to a learning model, as will be considered in this work. 

\subsection{LDPC code construction}
We now briefly describe LDPC code construction, given that this construction is known to have an impact on the code performance for data reconstruction, and may also influence the learning performance.  

The parity check matrix $H$ of an LDPC code can be equivalently described as a Tanner graph, which is bipartite in nature, connecting the $N$ source nodes (also called variable nodes) with the $M$ syndrome nodes (or check nodes). The degree of a given source node $n \in \llbracket 1,N \rrbracket$ is denoted $d_n$ and it is given by the number of syndrome nodes to which source node $n$ is connected in the Tanner graph. In the same way, the degree of a given syndrome node $m \in \llbracket 1,M \rrbracket$ is denoted $\delta_m$. 

An LDPC code is said to be regular if all source and syndrome node degrees are constant, that is $d_n = d_v$ for all $n  \in \llbracket 1,N \rrbracket $ and $\delta_m = \delta_c$ for all $m  \in \llbracket 1,M \rrbracket$. In this case, the source coding rate can be expressed as 
\begin{equation}
R = \frac{d_v}{\delta_c} .
\end{equation}
If degrees differ within source nodes and syndrome nodes, the code is said to be irregular. In this case, the source nodes degree distribution $\lambda(x)$ and syndrome nodes degree distribution $\rho(x)$ are expressed in polynomial forms as 
\begin{align}
\lambda(x)  = \sum_{i=1}^{d_{\text{max}}} \lambda_i x^{i-1}, ~~~ \rho(x) = \sum_{j=1}^{d_{\text{max}}} \rho_j x^{j-1},
\end{align}
where $d_{\text{max}}$ is the maximum degree, 
 $\lambda_i$ is the proportion of edges of the Tanner graph connected to a source node of degree $i$, and $\rho_j$ is the proportion of edges of the Tanner graph connected to a syndrome node of degree $j$. For irregular codes, the source coding rate can also be expressed in convenient form from the degrees as
\begin{equation}
R = \frac{\sum_{j} \rho_j/j}{\sum_{i} \lambda_i/i} . 
\end{equation}
It is well known that the code performance depends on its degree distributions, and that irregular LDPC codes usually perform better than their regular counterparts~\cite{chung2001design}. 
The degree distributions can be optimized with standard code design tools such as density evolution, see~\cite{chung2001design} for more details. 

Once the code degree distributions are fixed, the next step is to construct an LDPC parity check matrix $H$ that satisfies these distributions. In this work, we employ the standard Progressive Edge Growth (PEG) algorithm~\cite{hu2005regular} which performs this construction while aiming to minimize the amount of short cycles in the graph. The amount of short cycles is indeed known to have an important effect on the final code performance.  

The code design rules briefly described in this section were initially derived in the context of channel coding, where the objective is to correct the largest proportion of errors from the channel. These rules were also updated when considering LDPC codes for source coding targeting data reconstruction, see \emph{e.g.},~\cite{ye2019optimized}.  We leave for future works the investigation of optimal design rules for the case where the coding objective is now related to a specific learning task. However, in the numerical results section, we will empirically investigate the effect of code parameters such as rate and degree distribution onto the learning performance. 

\section{Image classification over compressed data}
This section introduces our methodology for image classification over JPEG compressed data. 
We first describe the conventional JPEG encoding baseline, and how we insert LDPC codes in the compression pipeline. We then present the considered GRU model for image classification. 
\begin{figure}[htbp]
\centering \includegraphics[width=1\linewidth]{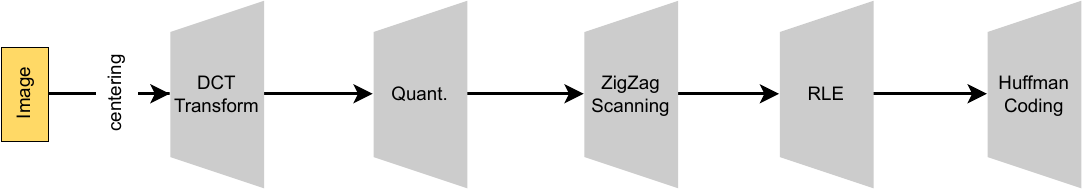}
\caption{Standard JPEG encoder Baseline}
\label{JPEG_pipeline}
\end{figure}

\label{preliminary}
\subsection{JPEG Encoding Baseline and the JPEG-LDPC Alternative}
The JPEG baseline described in Figure~\ref{JPEG_pipeline} consists of several steps designed to compress an image effectively. 
The resulting bitstream is made up of symbols obtained from a Huffman entropy encoding. The stages preceding Huffman encoding prepare the data for efficient probabilistic coding.

 \begin{figure*}[htbp]
\includegraphics[width=\linewidth]{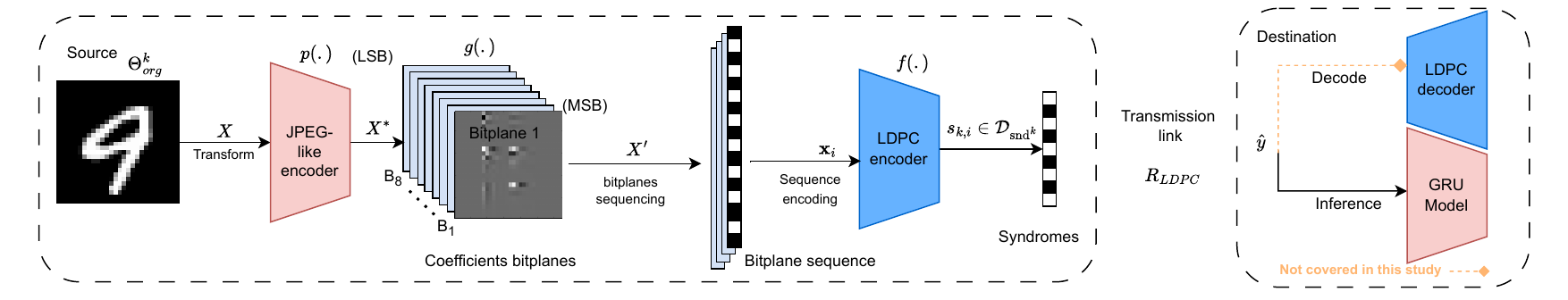}
\caption{\textbf{Second setup}: Syndromes of the DCT-LDPC coefficients bitplanes are fed as inputs of a GRU model for classification}
\label{setup2}
\end{figure*}

The first step is the DCT, which 
calculates the elementary frequency components of the original image. It also enables the concentration of spatial domain energy in the top-left zone, thereby preparing the data for subsequent steps in the JPEG process.
Next, quantization is performed by dividing the DCT coefficients by a fixed values table designed to align with the human psycho-visual system. The severity with which coefficients are sacrificed can be modulated by a quality factor, consequently increasing or decreasing the number of non-zero coefficients. The quantized coefficients are then reordered in a zigzag pattern to optimize data compaction for the Run Length Encoding (RLE) step. Finally, Huffman coding is employed based on the likelihood and category of each symbol's occurrence, specifically targeting the constant component coefficient at position $(0,0)$ of the block, referred to as discrete current (DC). The remaining $63$ coefficients are termed alternating components (AC). For conciseness, we do not describe in detail each of these steps which are standard in JPEG compression. In what is next, we focus on what is specific to our approach, that is replacing Huffman coding with LDPC codes.

\begin{figure}[htbp]
\centering \includegraphics[width=\linewidth]{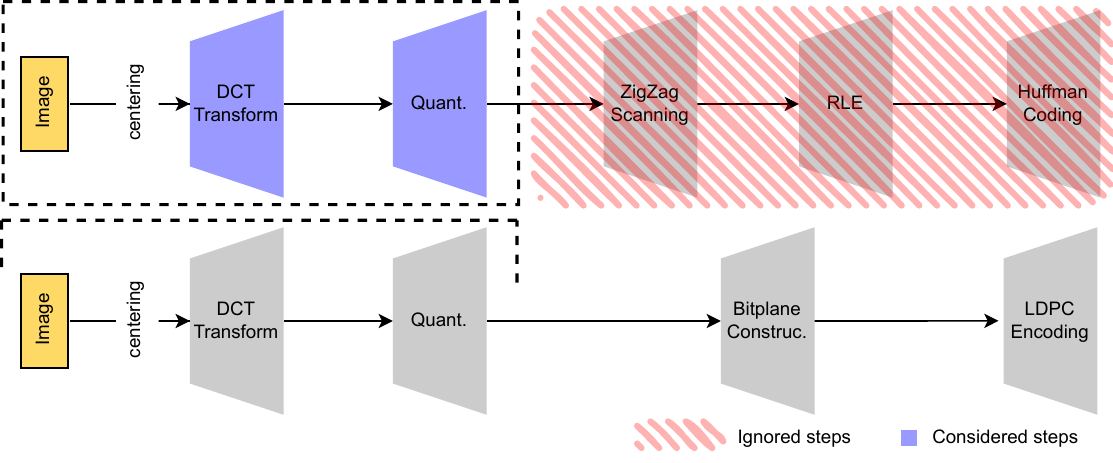}
\caption{Considered JPEG-LDPC encoder pipeline}
\label{pipeline_jl}
\end{figure}


\subsection{First Setup: Learning over LDPC-coded images}
\label{proposedmeth}

We consider an initial dataset of images $\Theta_{\text{org}}$ and investigate two encoding setups that we now describe. 

In the first setup, illustrated in Figure~\ref{setup1}, we consider source coding using only binary LDPC codes. We do not consider JPEG compression: for each image $\boldsymbol{I}_{\ell} \in \Theta_{\text{org}}$, the pixels are directly transformed into $K$ bitplanes and encoded using LDPC codes following the approach described in Section~\ref{sec:ldpc_codes}. The $K$ resulting syndromes $\mathbf{s}_{\ell,k}$ form new data $\mathbf{d}_{\ell}^{(1)} = \{ \mathbf{s}_{\ell,k} \}_{k \in \llbracket 0,K-1\rrbracket}$. We refer to the corresponding dataset as $\mathcal{D}_{\text{snd}}^{(1)}$, where the generation of element $\mathbf{d}_{\ell}^{(1)}$ of the dataset $\mathcal{D}_{\text{snd}}^{(1)}$ can be described as follows:
\begin{equation}\label{eq:enc1}
\mathbf{d}_{\ell}^{(1)} = f(g(\boldsymbol{I}_{\ell})) .  
\end{equation}
Here, the function $g(\cdot)$ represents the bitplane constructor, and the function $f(\cdot)$ is the LDPC source encoding according to Equation~\eqref{eq:ldpc_enc}. 

The objective of considering this setup is to investigate whether learning over LDPC source-coded bitplanes is efficient in the sense of preserving some initial common features of the images within the same class. Indeed, according to Equation~\eqref{eq:ldpc_enc}, LDPC codes achieve dimensionality reduction through a linear operation. We expect this linear operation to preserve distances between images in the compressed domain, as in compressed sensing~\cite{donoho2006compressed}. 

%


\subsection{Second Setup: Learning over LDPC-JPEG coded images}\label{sec:second_setup}
For the second setup, we consider JPEG compression and first apply the DCT transform followed by standard JPEG quantization. We do not consider Zig-Zag scanning nor RLE (since their primary role is preparing the data for Huffman coding) and directly transform the DCT coefficients into bitplanes. In this case, each element $\mathbf{d}_{\ell}^{(2)}$ of the  dataset  $\mathcal{D}_{\text{snd}}^{(2)}$ is obtained as
\begin{equation}
    \mathbf{d}_{\ell}^{(2)} = f(g(p(\boldsymbol{I}_{\ell}))) .  
\end{equation}
The function $p(\cdot)$ represents the JPEG-like operation, which carries out $8\times8$ DCT transform and quantization.  For the quantization step, we employ the standard JPEG quantization matrix. The functions $f$ and $g$ are the same as in Equation~\eqref{eq:enc1}. In the numerical results section, we will evaluate the classification performance of this coding scheme with different JPEG parameters, such as the quality factors. 

\subsection{Classification Model: Gated Recurrent Units}
 We now introduce the classification model used in the two considered setups for learning over compressed data.
We consider a classification task with $C$ classes, and we use the conventional cross-entropy as loss function. The GRU model~\cite{cho2014properties} we consider is a simplified variant of the Long Short-Term Memory (LSTM)  framework~\cite{greff2016lstm}, combining the forget and input gates into a single update gate and introducing a reset gate. This reduction in complexity is achieved without substantially reducing performance. Unlike LSTMs, GRUs manage the flow of information without a separate memory cell, simplifying the processing of temporal sequences.

\begin{figure*}[htbp]\centering \includegraphics[width=1\linewidth]{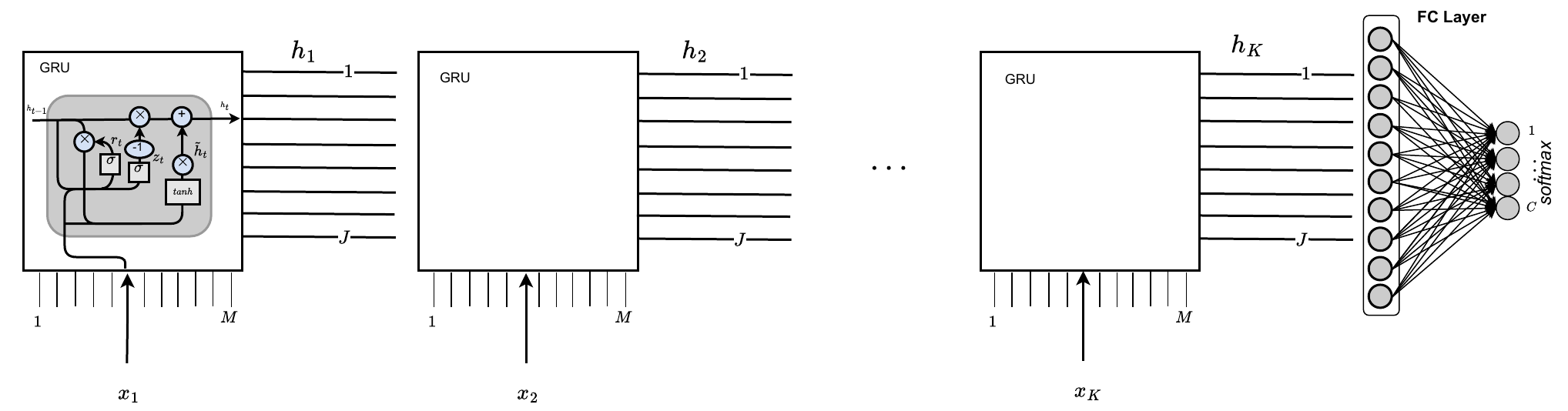}
\caption{Considered classification model for classification using $J$ units of one GRU layer.}
\label{model}
\end{figure*}




The considered GRU architecture is shown in Figure~\ref{model}. 
We use a GRU with $J$ units and $K$ time steps, where $K$ is the number of bitplanes. Each time step $k \in \llbracket 1,K\rrbracket$ takes as input bitplane $\mathbf{x}_k$. The hidden state at time $k$ is denoted by the vector $\mathbf{h}_k \in \mathbb{R}^{J}$, whose $j$-th component is $h_{k,j}$ for $j \in \llbracket 1,J\rrbracket$. 

In the GRU model, the hidden state is updated by combining the previous state $\mathbf{h}_{k-1}$ with a candidate hidden state $\tilde{\mathbf{h}}_{k}$, using the following equation 
\begin{equation}\label{eq:gru_update}
\mathbf{h}_{k} = (1 - \mathbf{z}_{k}) \odot \mathbf{h}_{k-1} + \mathbf{z}_{k} \odot \tilde{\mathbf{h}}_{k},
\end{equation}
where $\odot$ denotes pointwise multiplication, and $\mathbf{z}_{k}$ is the update gate  at time $k$.
The update gate vector $\mathbf{z}_k \in \mathbb{R}^{J}$ controls how much of the previous state is retained. Its expression is:
\begin{equation}
\mathbf{z}_{k} = \sigma\left(\mathbf{W}_{z}\mathbf{x}_{k} + \mathbf{U}_{z}\mathbf{h}_{k-1}\right), 
\end{equation}
where the $k$-th syndrom $\mathbf{x}_{k} \in \mathbb{R}^{M}$ is the input at time $k$, and $\sigma(\cdot)$ is the element-wise sigmoid function. The weight matrices $\mathbf{W}_{z} \in \mathbb{R}^{J \times M}$ 
and $\mathbf{U}_{z} \in \mathbb{R}^{J \times J}$ respectively map the input and the previous hidden state to the update gate.
The candidate hidden state $\tilde{\mathbf{h}}_{k}$ is then computed as:
\begin{equation}
\tilde{\mathbf{h}}_{k} = \tanh\left(\mathbf{W}\mathbf{x}_{k} + \mathbf{U}(\mathbf{r}_{k}\odot\mathbf{h}_{k-1})\right),
\end{equation}
where $\mathbf{r}_{k} \in \mathbb{R}^{J}$. The matrix $\mathbf{W}$ is of size $J\times M$, and the matrix $\mathbf{U}$ is of size $J\times J$.  The reset gate determines how much of the previous hidden state is used when computing the candidate state:
\begin{equation}
\mathbf{r}_{k} = \sigma\left(\mathbf{W}_{r}\mathbf{x}_{k} + \mathbf{U}_{r}\mathbf{h}_{k-1}\right).
\end{equation}
When $r_{k,j}=0$, the $j$-th unit at time $k$  is effectively reset, ignoring previous information. The weight matrix $\mathbf{W}_r$ is of size $J\times M$, and the weight matrix $\mathbf{U}_r$ is of size $J\times J$.
By controlling the contributions of the previous hidden states and the current input, the GRU can capture both short-term and long-term dependencies. 


The last GRU layer produces a hidden state vector $\mathbf{h}_{K} \in \mathbb{R}^{J}$ at the final time step $K$. 
Therefore, a fully connected (FC) layer is next applied to $\mathbf{h}_{K}$. Given that the classification task involves $C$ classes, the FC layer maps $\mathbf{h}_{K} \in \mathbb{R}^{J}$ to a $C$-dimensional output $\mathbf{o} \in \mathbb{R}^{C}$:
\begin{equation}
\mathbf{o} = \mathbf{W}_{\text{FC}}\mathbf{h}_{K} + \mathbf{b}_{\text{FC}},
\end{equation}
where $\mathbf{W}_{\text{FC}} \in \mathbb{R}^{C \times J}$ and $\mathbf{b}_{\text{FC}} \in \mathbb{R}^{C}$ are the parameters of the fully connected layer. The output $\mathbf{o}$ is then passed through a softmax function to obtain the probability of each class.

It is worth mentioning that a Stacked GRU model was considered in~\cite{bennatan2018deep}
 for the maximum-likekihood decoding of short linear block codes. Our work differs in the sense that the learning objective is not the same (decoding of a full vector after passing through a noisy channel in~\cite{bennatan2018deep}, versus classification over compressed data in this paper). Also, the GRU architecture differs (Stacked GRU with several layers with the same input at each timestep in~\cite{bennatan2018deep}, only one layer and one different bitplane as input at each timestep in our work). Finally,the work in~\cite{bennatan2018deep} is only suitable for short codes ($128$ bits at maximum) given that the maximum-likelihood decoding problem is much more complex, while in our case, we apply our model applies to longer codes, for instance considering $N=1024$ bits. However, both the results of~\cite{bennatan2018deep} and our work confirm that the GRU model is especially relevant when considering coding from linear block codes.  


\section{Data and Experiments}\label{sec:data_exp}
We now describe the image datasets as well as the experimental setups we use to evaluate the classification performance of the two considered setups.

\subsection{Datasets}
In our experiments, we consider four datasets: MNIST~\cite{lecun1998gradient}, Fashion-MNIST~\cite{xiao2017fashion}, CIFAR-10~\cite{krizhevsky2009learning}, and Y-CIFAR-10, derived from CIFAR-10. 
For MNIST and Fashion-MNIST, images are converted to grayscale and processed as outlined in subsection~\ref{cod_par}. This provides $K=8$ bitplanes per image.  
To construct the Y-CIFAR-10 dataset, CIFAR-10 images are converted to YCrCb color space. 
Only the Y luminance channel is considered for coding, as it is known to encapsulate the majority of the information of the original RGB images. This also outputs $K=8$ bitplanes for the CIFAR dataset. 
Finally, we use the NTSC color space (luminance and chrominance) to represent the  CIFAR-10 dataset. Our empirical tests indeed showed an improvement of $1\%$ to $2\%$ compared to the RGB representation. The NTSC color space is obtained using the following equation \cite{ITU_R_BT1700}
\begin{equation}
\begin{bmatrix}
Y \\
I \\
Q
\end{bmatrix}
=
\begin{bmatrix}
0.299 & 0.587 & 0.114 \\
0.596 & -0.274 & -0.322 \\
0.211 & -0.523 & 0.312
\end{bmatrix}
\begin{bmatrix}
R \\
G \\
B
\end{bmatrix}
\end{equation}
Here, the three components of this color space are used. No downscaling is performed for the I and Q components. This CIFAR-10 dataset is used when evaluating the method for a larger number of bitplanes, $K=24$. 






\subsection{LDPC code parameters}
\label{cod_par}



In order to maintain consistency in the evaluation, we always consider a code length $N=1024$, which corresponds to images with $1024$ pixels. Consequently, the sizes of the images of the four datasets are resized from $28 \times 28$ to $32 \times 32$ by padding zeros at the end of the columns and rows. This adjustment also facilitates the application of the $8\times8$ DCT transform. 
In our experiments, we further consider both regular and irregular LDPC codes, with different rates $1/4$, $1/2$, $3/4$. The considered regular and irregular codes are shown in Table~\ref{tab:code_param}. 

\begin{table}[htbp]
    \caption{Regular and irregular LDPC codes parameters}

    \centering
    \begin{tabular}{p{0.06\columnwidth}|p{0.15\columnwidth}|p{0.32\columnwidth}|p{0.25\columnwidth}}
    \toprule
   \multicolumn{4}{c}{Regular codes} \\
     \toprule
        Rate & size of $H$ & $d_v$ & $\delta_c$ \\ \midrule
        $1/4$ & $256 \times 1024$   & $3$  & $12$ \\ 
        $1/2$ & $512 \times 1024$ & $3$ & $6$ \\ 
        $3/5$ & $615 \times 1024$ & $3$ & $5$ \\ 
        $3/4$ & $768 \times 1024$ & $3$ & $4$ \\ 
         \toprule   
         \multicolumn{4}{c}{Irregular codes} \\
     \toprule
        Rate & size of $H$ & $\lambda(x)$ & $\rho(x)$ \\ \midrule
        $1/4$ & $256 \times 1024$   & $0.25x + 0.75x^2$  & $11x^{10} $ \\ 
        $1/2$ & $512 \times 1024$   & $0.5x + 0.25x^3 + 0.25x^4$  & $0.5x^5 + 0.5x^6 $ \\ 
        $3/4$ & $768 \times 1024$ & $0.75x + 0.25x^3$ &  $0.67x^2 + 0.33x^3$\\ 
         \bottomrule   
    \end{tabular}
    \label{tab:code_param}
\end{table} 


%
%
%


\subsection{Model parameters}


In our experiments, the input size of the GRU model is set according to the compression rate used in the LDPC coding step. Firstly, when no coding is performed, the input size is set to $N=1024$, and the GRU model is applied over the original bitplanes. 
When LDPC encoding is applied, the input size is set to $M$, with $M=256$ for rate $1/4$, $M=512$ for rate $1/2$, and $M=786$ for rate $3/4$.
In every case, the number of time-steps is given by the number of bitplanes $K$, and different values of $K$ ranging between $1$ and $24$ will be considered in our experimentation. The other model hyper-parameters are provided in Table~\ref{tab:learning_param}.

For model training, the datasets are divided into training/validation sets with $60000$ samples for training and $10000$ samples for validation for MNIST and Fashion-MNIST, and $50000$ samples for training and $10000$ samples for validation for Y-CIFAR-10 and CIFAR-10. 

\begin{table}[htbp]
    \caption{Hyper-parameters for the learning model}

    \centering
    \begin{tabular}{p{0.24\columnwidth}|p{0.15\columnwidth}|p{0.25\columnwidth}|p{0.15\columnwidth}}
     \toprule
        \textbf{parameter} & \textbf{Value} & \textbf{parameter} & \textbf{Value} \\ \midrule
        Learning rate &  0.001 & Optimizer & Adam \cite{kingma2014adam} \\  
        No. epochs & 30 & L2 Regularization & 0.002  \\  
        No. GRU Units (J) & [12, 32, 50] & Activation function & Softmax \\ 
        Batch size & 64 & GRU activation& Tanh \\ 
         \bottomrule
        
    \end{tabular}
    \label{tab:learning_param}
\end{table}

\subsection{Evaluation metrics}
We now describe the specific metrics that will be considered in our performance evaluation. Of course, conventional accuracy will be used to evaluate the classification performance. 
\subsubsection{Approximate entropy}
We adopt the metric proposed in \cite{piau2023predicting} to measure the level of chaos introduced into the data by several entropy-coding methods (Huffman, arithmetic, LDPC). This metric uses the Approximate Entropy (ApEn) and it is defined for a given image $\mathbf{I}$ as 
\begin{equation}
    D_{f}(\mathbf{I}) = \frac{\sum_{k=0}^{K-1} \text{ApEn}_{m,r}(f(g_k(\mathbf{I})))}{\sum_{k=0}^{K-1} \text{ApEn}_{m,r}(g_k(\mathbf{I}))} .
    \label{apen}
\end{equation}
In this equation, given that \(\mathbf{I}\) represents the original image, the function denoted $g_{k}:\{0, 1\}^* \rightarrow \{0, 1\}^*$ is utilized to extract the bitplanes of this image, and $f$ is the LDPC encoding. Furthermore, the quantity $\text{ApEn}_{m,r}$ is calculated based on the window size specified by $m$ and a threshold determined by $r$. Here, we set $m=2$ and $r=0.2$. 
This metric approximates entropy and it has been widely used for instance in time series to measure the predictability of the data through entropy assessment~\cite{pincus1991approximate}. 

%
%

\subsubsection{Rate gain}
We also measure the rate gain when considering LDPC bitplanes coding using the following equation:
\begin{equation}\label{eq:rate_gain}
     \Gamma = \frac{R}{N_{\text{bp}}},
\end{equation}
where $R$ is the LDPC source coding rate, and $N_{bp}$ is calculated as the number $K$ of considered bitplanes divided by the maximum number of bitplanes $\eta_{bp}$, ie: $N_{bp}=K/\eta_{bp}$.  

\begin{figure*}[htbp]
  \begin{center}
  \includegraphics[width=1\linewidth]{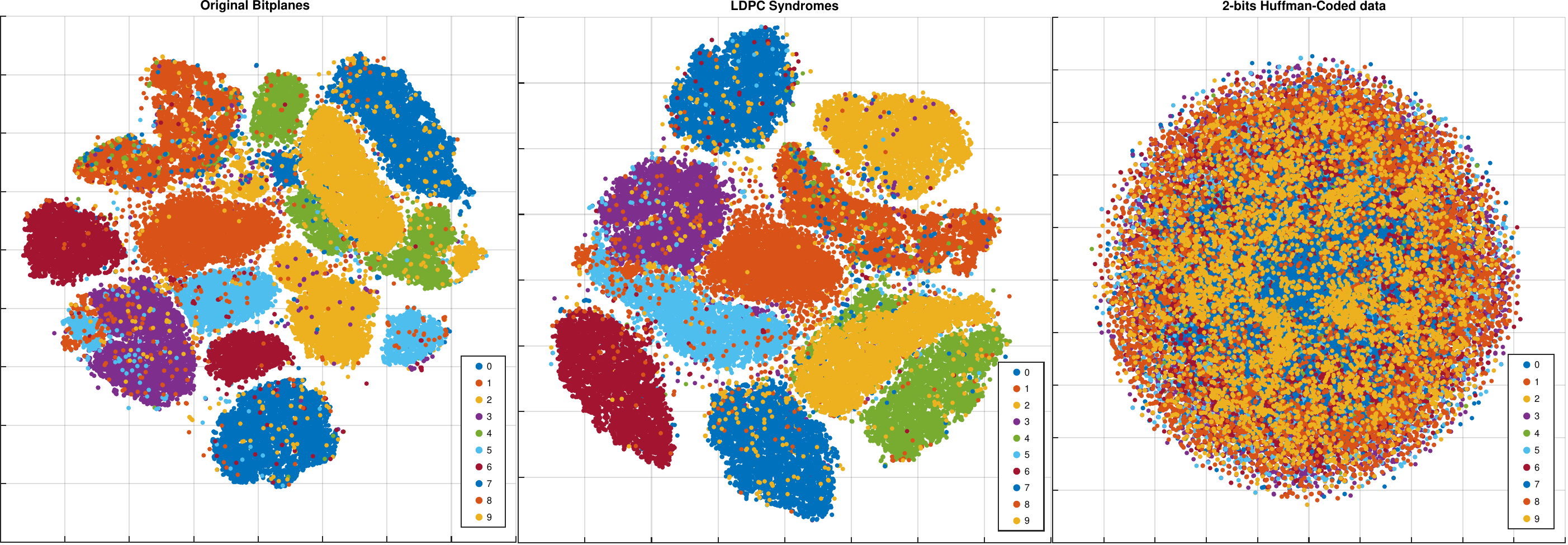}
  \caption{t-SNE analysis for class separability on the MSB ($1^{st}$ bitplane) of MNIST dataset (60000 samples) in the original non-coded bitplanes (left), 1/2 regular LDPC Syndromes (Middle), and when coded with 2 bits Huffman (right).}\label{chaos_tsne}
  \end{center}
\end{figure*}
\section{Numerical Results and Discussion}
This section presents our numerical results for the classification performance of the proposed approach with LDPC codes. It starts with an analysis of the ability of LDPC codes to support classification, using a t-SNR analysis as well as the notion of Approximate Entropy. Then, the section provides a comparison in terms of accuracy with state-of-the-art approaches. Finally, it evaluates the impact of various parameters related to LDPC codes and JPEG compression. 

\label{sec:results_disc}
\subsection{Quantifying Chaos introduced by entropy-coding}
Our first analysis focuses on evaluating the impact of entropy coding on the intrinsic structure of the data. For this purpose, we consider both the t-SNE Analysis and the Approximate Entropy metric given in equation~\eqref{apen}.

\subsubsection{t-SNE Analysis on MNIST}
t-SNE (t-distributed stochastic neighbor embedding) \cite{van2008visualizing} is a technique extensively used for dimensionality reduction. This technique maps high-dimensional datasets into two-dimensional spaces, revealing the underlying class structure. 
Here, we utilize the t-SNE technique to visualize the degree of separation between classes. We use only the first MSB bitplane of the entire MNIST dataset ($60000$ samples) and consider a regular LDPC matrix with rate 1/2. We make a comparison with the $2$-bit Huffman coding method implemented in \cite{remy1}. 
%

%
%
Figure (\ref{chaos_tsne}) presents the t-SNE visualizations of the MNIST dataset, evaluated for uncoded data, for data after LDPC coding, and for data after Huffman coding. We see that the uncoded dataset exhibits distinct, well-separated clusters characterized by high intra-class variability. Applying LDPC codes reduces intra-class variability, leading to more compact clusters, attributed to the limited number of syndromes. However, it still shows adequate inter-class separability. 
On the opposite, Huffman coding markedly diminishes class separability, producing overlapping clusters. We conclude that LDPC codes may be more appropriate than Huffman coding for classification in compressed-domain. 

\subsubsection{Approximate Entropy-based chaos measurement on CIFAF-10 }
\begin{figure}[htbp]
  \begin{center}
  \includegraphics[width=1\columnwidth]{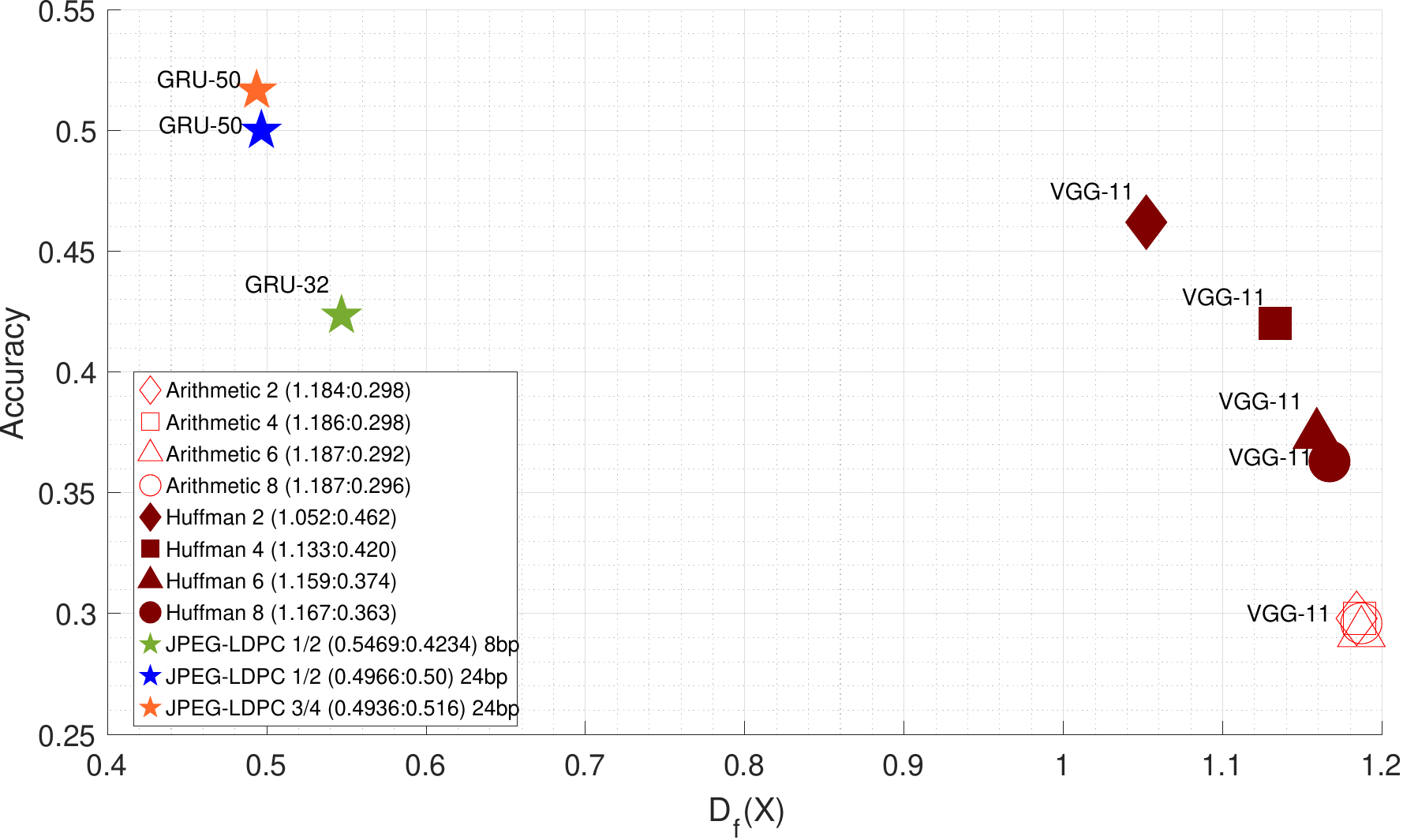}
  \caption{Chaos measurement using $D_f(X)$ against accuracy for YCIFAR-10 and CIFAR-10 datasets}
  \label{chaos_f}
  \end{center}
\end{figure}

Figure (\ref{chaos_f}) shows the classification accuracy with respect to $D_f(\mathbf{I})$ for CIFAR-10 and YCIFAR-10 datasets. Here we consider our second setup described in Section~\ref{sec:second_setup} (JPEG-LDPC), using the regular LDPC code with rates $1/2$ and $3/4$. We consider a $32$-units ($52.6$k parameters) GRU model with $8$ bit planes, and a $50$-units ($85$k parameters) GRU model with $24$ bitplanes. The comparison is made against the methods reported in~\cite{piau2023predicting} that produce Huffman symbols and Arithmetic symbols for groups of bits ($2$ bits, $4$ bits, and $8$ bits, respectively). The work in~\cite{piau2023predicting} uses UVGG11~\cite{simonyan2014very} (Unidimensional VGG11) as a deep learning model to classify the compressed bitstreams ($\sim 35$M parameters). 
We observe that the $D_{f}(\mathbf{\mathbf{I}})$ metric has a significantly lower value for the LDPC-GRU pair. 
We observe that the GRU-LDPC model with $32$ units shows similar accuracy as for the VGG-$11$ model applied on Huffman coded data.  In addition, the larger GRU model with $50$ units greatly improves the accuracy, in accordance with the lower value of $D_f$. 
This analysis shows that the LDPC dataset is markedly more structured than the Huffman-coded and arithmetic-coded ones. Therefore, considering suitable Deep Learning models provides higher accuracy on the LDPC dataset. 

\subsection{Accuracy comparison}

\begin{table*}[htbp]
\centering

\caption{Classification accuracy of the GRU on LDPC coded data compared to the state-of-the-art for multiple datasets.}
\begin{tabular}{ll|cc|ccc|ccccc}

\toprule

\label{tab:performance_metrics}

\textbf{Dataset} & \textbf{Model} & \multicolumn{2}{c}{\textbf{No coding}} & \multicolumn{3}{c}{\textbf{Coding on Orig. (Setup1)}} & \multicolumn{5}{c}{\textbf{Coding on JPEG (Setup2)}} \\ 
\cmidrule(lr){1-1} \cmidrule(lr){2-2} \cmidrule(lr){3-4} \cmidrule(lr){5-7} \cmidrule(lr){8-12} 
 &  & \textbf{None} & \textbf{\makecell{None\\MSB}} & \textbf{Huff\cite{remy1}} & \textbf{Arith\cite{remy1}} & \textbf{LDPC} & \textbf{JPEG\cite{remy1}} & \textbf{\makecell{DCT\\-tr.\cite{fu2016using}}} & \textbf{\makecell{J-L\\8bp}} & \textbf{\makecell{J-L\\MSB}} &\textbf{\makecell{J-L\\MSB+1bp}}\\ 
\midrule
\multirow{2}{*}{\textbf{\textit{MNIST}}} &\textbf{GRU12(proposed)} & 0.9439 & 0.8842 & 0.6790 & 0.5086 & \textbf{0.8192}  & - & - & \textbf{0.9060} & 0.6548&0.8791\\
                       &\textbf{GRU32(proposed)} & 0.9799 & 0.9154 & 0.7563 & 0.5370 & \textbf{0.8556} & - & - & \textbf{0.9237} & 0.6843&0.8849\\
                              \cmidrule(lr){2-12}
                       & \textbf{UVGG11 \cite{remy1}} & 0.9891 & - & 0.8323 & 0.6313 & - & -  & - & - & -&-\\
                       & \textbf{URESNET18 \cite{remy1}} & 0.9875 & - & 0.7450 & 0.5949 & - & -  & - & - & -&-\\
                              \cmidrule(lr){2-12}
                       & \textbf{FullyConn \cite{fu2016using}} & 0.9200 & - & - & - & -  & - & $0.9000$ & - & -&-\\
\midrule
\multirow{2}{*}{ \makecell{\textbf{\textit{Fashion}} \\ \textbf{\textit{-MNIST}}}} &\textbf{GRU12} & 0.8616 & 0.8052 & - & - & \textbf{0.8166}  & - & - & \textbf{0.8332} & 0.5222&0.8325\\
                               &\textbf{GRU32} & 0.8750 & 0.8314 & - & - & \textbf{0.8306}  & - & - & \textbf{0.8434} & 0.5395&0.8414\\
                                      \cmidrule(lr){2-12}
                               & \textbf{UVGG11 \cite{remy1}} & 0.9018 & - & 0.7634 & 0.6898 & - & -  & - & - & -&-\\
                               & \textbf{URESNET18 \cite{remy1}} & 0.8497 & - & 0.6862 & 0.6116 & - & -  & - & - & -&-\\
\midrule
\multirow{2}{*}{\makecell{\textbf{\textit{YCIFAR}}\\\textbf{\textit{-10}}}} &\textbf{GRU12} & 0.3127 & 0.3249 & 0.2374 & - & \textbf{0.4070}  & - & - & \textbf{0.4234} & 0.1350&0.3537\\
                           &\textbf{GRU32} & 0.3596 & 0.3560 & 0.2400 & - & \textbf{0.4171} & - & - & \textbf{0.4316} & 0.1403&0.3544\\
                                  \cmidrule(lr){2-12}
                           & \textbf{UVGG11 \cite{remy1}} & 0.5657 & - & 0.3606 & 0.2976 & -  & 0.3245 & - & - & -&-\\
                           & \textbf{URESNET18 \cite{remy1}} & 0.3836 & - & 0.2591 & 0.2432 & -  & - & - & - & -&-\\
                                  \cmidrule(lr){2-12}
                           & \textbf{FullyConn \cite{fu2016using}} & 0.3800 & - & - & - & -  & - &  0.3000$ $ & - & -&-\\
\bottomrule
\multicolumn{12}{l}{\scriptsize *J-L: JPEG-LDPC, MSB+1bp: Sign bitplane of the DCT coefficients + the next bitplane after the sign bitplane} 
\end{tabular}

\end{table*}

Table~\ref{tab:performance_metrics} shows the classification accuracy over LDPC-coded images under different conditions, including: (i) without coding (for reference), (ii) by applying LDPC coding directly to the original image bitplanes (Setup $1$), (iii) with JPEG-LDPC compression (Setup $2$). The results are compared to those of the Huffman and Arithmetic coding methods from ~\cite{remy1}, and to the ``truncated DCT" technique of~\cite{fu2016using}. 

For Setup $1$, the training is applied over the $K=8$ bitplanes. For setup 2, we report the impact of the number of bitplanes on the model performance, when learning over $1$ bitplane (J-L MSB), $2$ bitplanes (J-L MSB + $1$ bp), and $8$ bitplanes (J-L 8bp). It is worth noting that the results of~\cite{piau2023predicting} for Huffman and Arithmetic coding only considered learning over the $8$ bitplanes.
Interestingly, we remark that in Setup $1$, our approach significantly outperforms state-of-the-art methods in terms of classification accuracy.  We observe a performance improvement of about $15\%$ for CIFAR-10 and $10\%$ for Fashion-MNIST and MNIST, compared to the best results of the studies mentioned above.

Table~\ref{tab:performance_metrics} also shows that Setup 2, which combines DCT and quantization with LDPC coding, surpasses the performance of Setup 1, which is solely based on LDPC coding. This superiority most probably comes from the DCT and its ability to concentrate the signal energy on certain coefficients. 
Furthermore, examining the impact of utilizing a limited number of bitplanes (columns J-L-MSB and J-L-MSB+bp) reveals interesting outcomes. Although learning from only the MSB bitplane $(K=1)$ degrades performance, learning from the MSB bitplane and one additional bitplane $(K=2)$ yields results nearly equivalent to using all bitplanes. This finding shows the feasibility of compressing data by considering a reduced number of biplanes without compromising the learning performance.



\subsection{Impact of LDPC coding rate}
\begin{figure}[htbp]
    \centering
    \includegraphics[width=1\columnwidth]{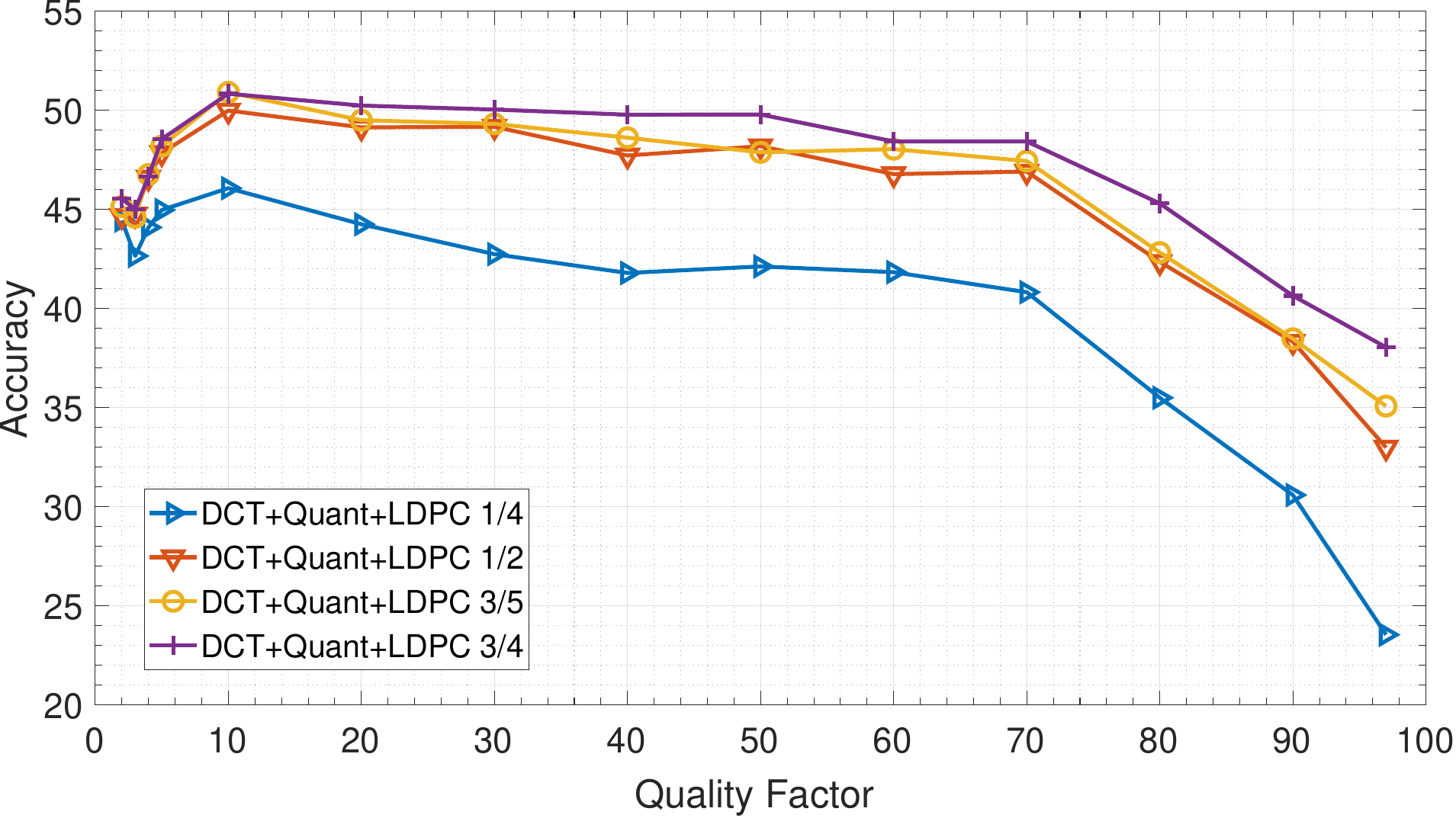}
    \caption{Impact of rate on regular LDPC codes for different JPEG Quality Factor}
    \label{fig:qf_rate_reg}
\end{figure}

From now on, we evaluate the impact of various parameters on the classification accuracy, starting with the LDPC source coding rate. 
Figure~\ref{fig:qf_rate_reg} provides the classification accuracy of the CIFAR-10 dataset as a function of JPEG QF for the four regular LDPC codes provided in Section~\ref{cod_par} using $24$ bitplanes. 
The curves show that the rate of LDPC coding significantly impacts the accuracy of the classification. Higher rates (\emph{i.e.}, lower compression ratios) consistently achieve better accuracy across all quality factors. This indicates that like for data compression, 
there is a trade-off between the LDPC code rate and the classification performance. 



\subsection{Regular vs irregular LDPC codes}
\begin{figure}[htbp]
    \centering
    \includegraphics[width=1\columnwidth]{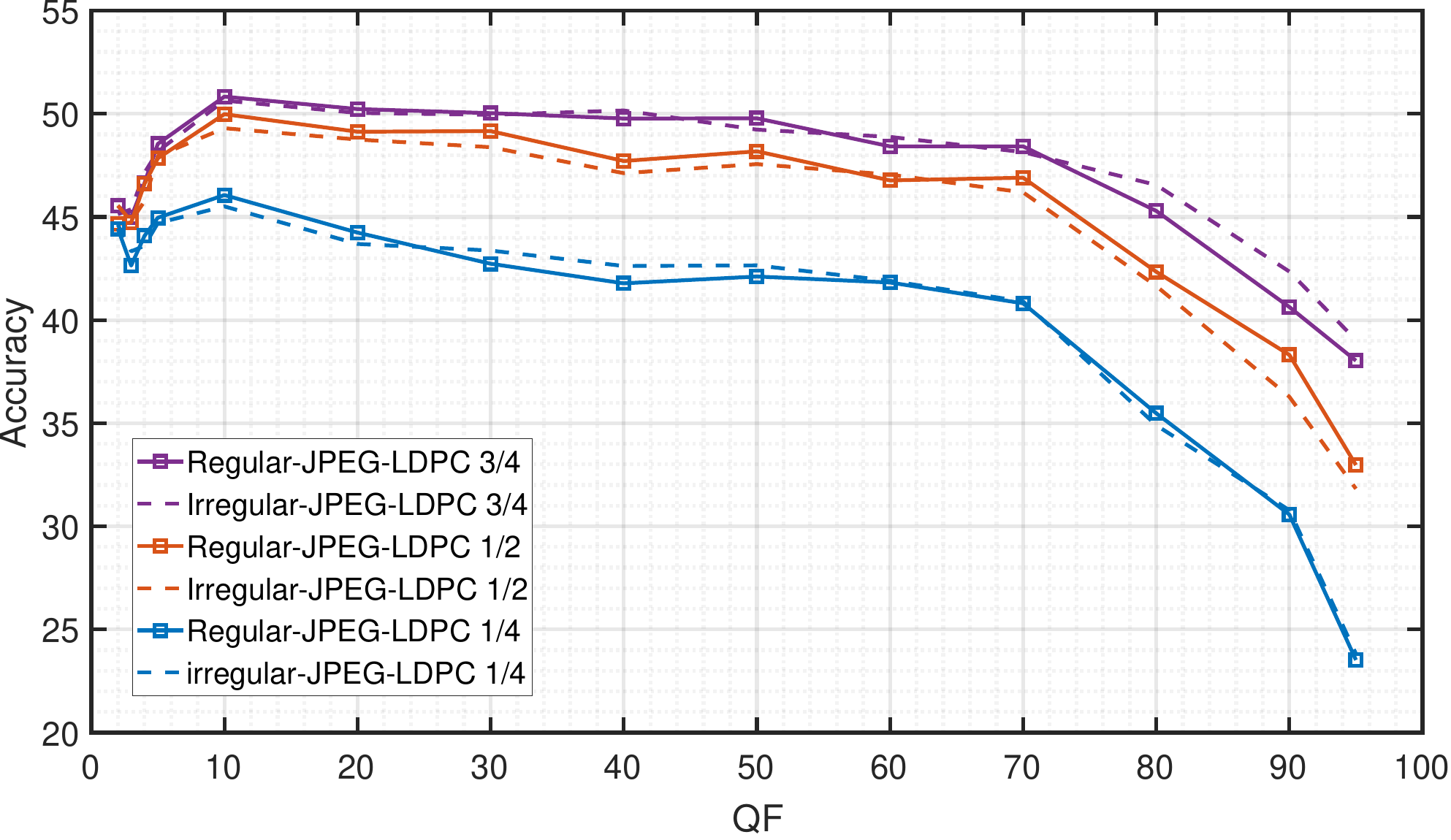}
    \caption{Impact of LDPC code type on learning performances (Regular vs. Irregular)}
    \label{fig:qf_rate_reg_irreg}
\end{figure}

In terms of data reconstruction, irregular LDPC codes have demonstrated superior performance compared to regular codes. 
Here, we evaluate the accuracy of our GRU model when images are compressed using either regular or irregular LDPC codes at different coding rates. We consider the CIFAR-10 dataset with $24$ bitplanes and $J=50$ units of the GRU model. 
Figure~\ref{fig:qf_rate_reg_irreg} shows the accuracy obtained from learning on the LDPC syndromes of Setup 2, considering multiple JPEG quality factors. The curves show that the difference in accuracy between regular and irregular LDPC codes is very small. However, irregular codes exhibit a slight improvement in accuracy, particularly at higher QFs. This trend is consistently observed across different compression rates, including low rates ($R=3/4$), moderate rates ($R=1/2$), and high rates ($R=1/4$). Note that the considered irregular codes were constructed following code design rules adapted to data reconstruction. Therefore, the investigation of code design rules specifically adapted to the learning problem appears to be a relevant perspective for future works.  

\subsection{Impact of quality factors}
We again consider the CIFAR-10 dataset with $24$ bitplanes. 
Figure~\ref{fig:qf_rate_reg} shows that the classification accuracy is almost constant for JPEG QF factors between $10$ and $60$. 
However, we then observe that, as QFs become high, there is a decrease in accuracy across all code rates. Initially, what was expected was that higher QF would yield more detailed and, consequently, more informative data. Nonetheless, at higher QF, the data appears to become excessively detailed or complex in a way that is not advantageous when encoded using LDPC. 
These results are also confirmed through the results presented in Figure~\ref{fig:prun} which evaluate the impact of truncation. 


\subsection{Impact of Pruning DCT coefficients on Learning:}

\begin{figure}
    \centering
    \includegraphics[width=1\columnwidth]{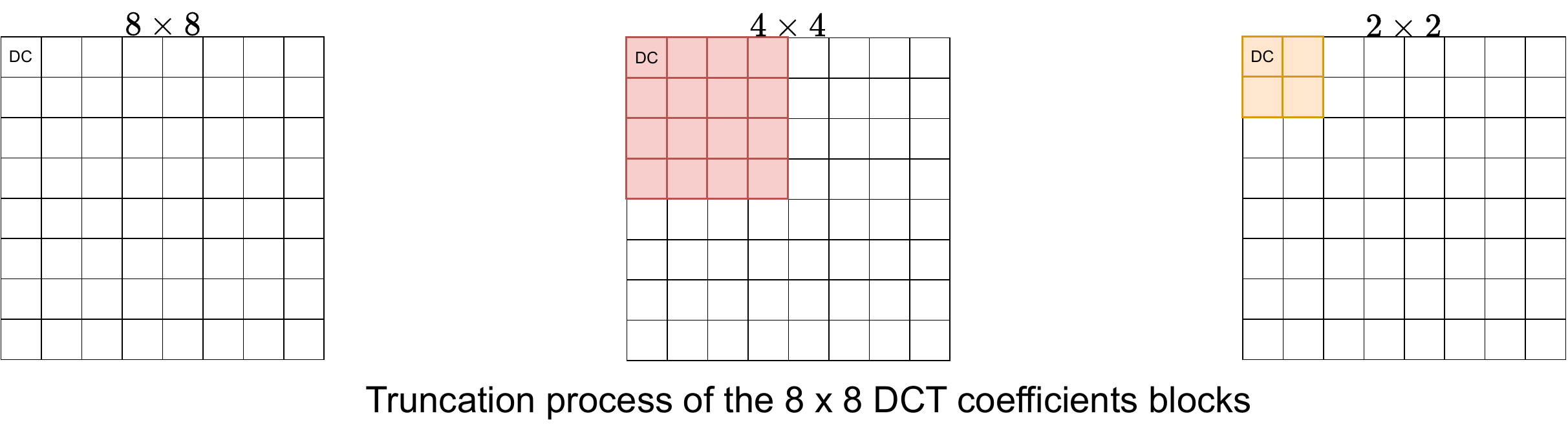}
    \caption{The truncation aims to take only a sub-block from the whole block of the DCT coefficients}
    \label{fig:prun2}
\end{figure}
\begin{figure}
    \centering
    \includegraphics[width=1\columnwidth]{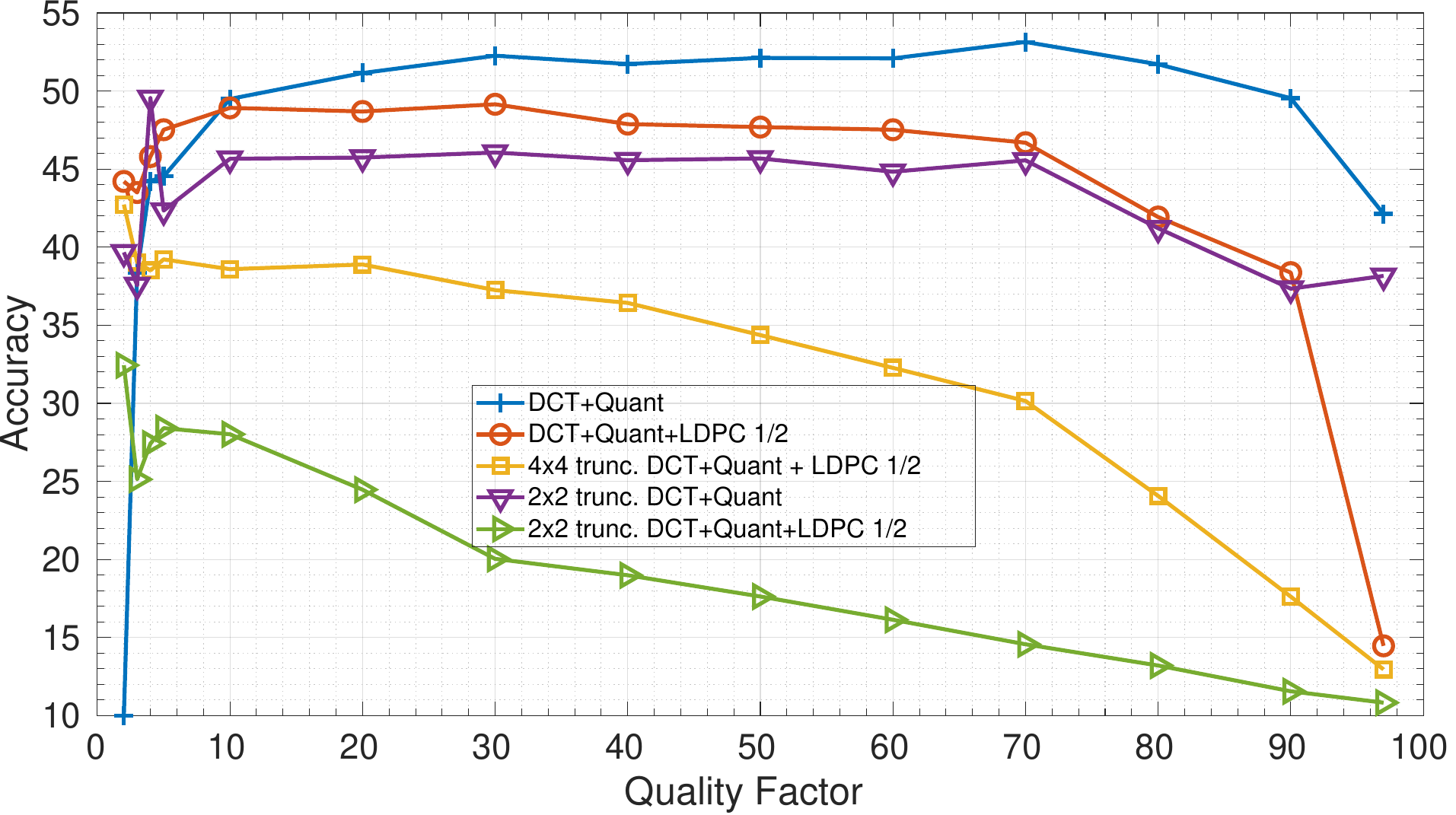}
    \caption{Impact of truncation, code length, and quality factor on 1/2 regular LDPC on colored CIFAR-10}
    \label{fig:prun}
\end{figure}

We here consider a truncation process shown in Figure (\ref{fig:prun2}), where the coefficients of subblocks of size either $4\times4$ or $2\times2$ are considered for the LDPC coding.
Figure (\ref{fig:prun}) illustrates the classification accuracy for different JPEG QFs for five methods:
\begin{enumerate}
    \item \textit{DCT + Quantization.}
    \item \textit{DCT + Quantization + LDPC $1/2$.}
    \item \textit{$4\times4$ truncated DCT + Quantization + LDPC $1/2$.}
    \item \textit{$2\times2$ truncated DCT + Quantization.}
    \item \textit{$2\times2$ truncated DCT + Quantization + LDPC $1/2$.}
\end{enumerate}

The DCT + quantization, employed as a baseline, provides an accuracy of approximately $40\%$ at a QF of $4$, and of $54\%$ at a higher QF of $97$.
When using the (DCT + Quant + LDPC 1/2) method, which means a reduction in data size of $50\%$, a comparable accuracy is obtained, reaching up to $49\%$ at $QF = 30$, demonstrating that a gain in data size by $50\%$ comes with a drop of only $5\%$ in accuracy. This again confirms that LDPC entropy-coding enables substantial data compression with negligible effects on learning accuracy. 

Next, when $4\times4$ truncation is applied along with LDPC coding, the accuracy is lower compared to the non-truncated LDPC method. This indicates that while truncation reduces the amount of data, LDPC coding helps preserve a reasonable level of accuracy. On the other hand, considering $2\times 2$ truncation with LDPC codes shows a severe performance degradation, due to the important loss of information in this case. 

\subsection{Rate gain}

\begin{figure}
    \centering
    \includegraphics[width=1\columnwidth]{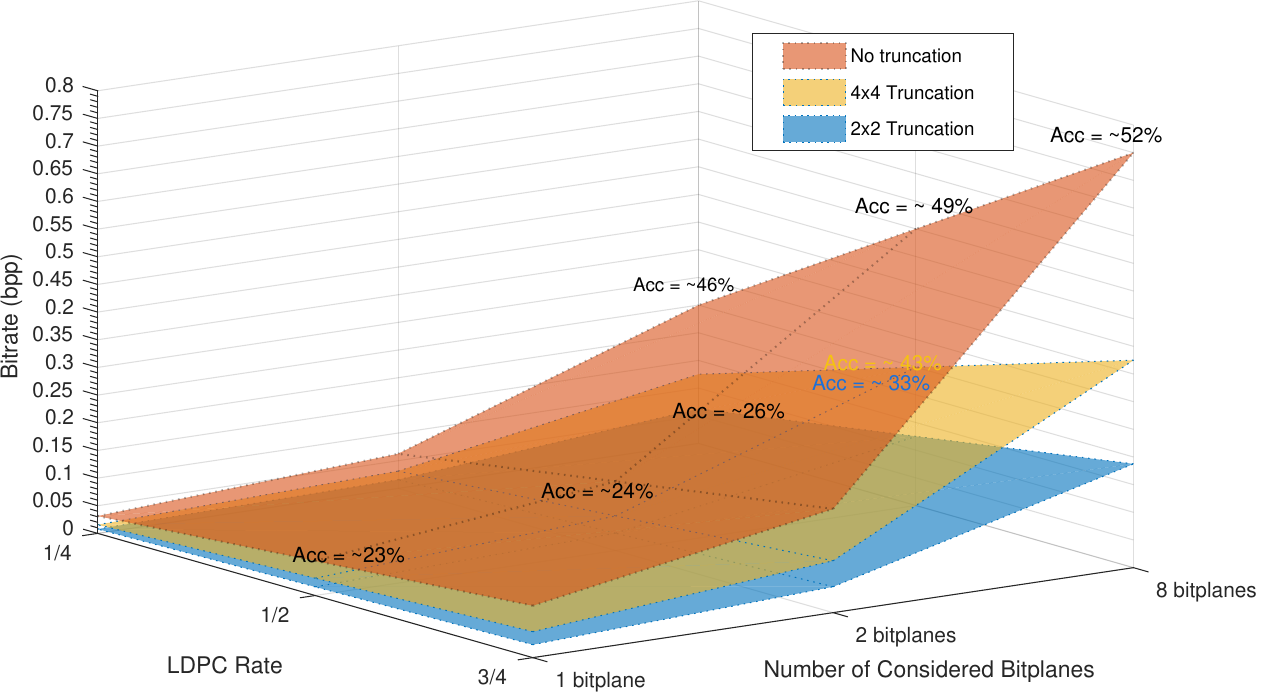}
    \caption{Rate gain expressed in bits per pixel with the corresponding achieved accuracy according to the number of retained bitplanes and the DCT truncation level for YCIFAR-10 Dataset.}
    \label{fig:rate_gain}
\end{figure}

Here, we consider the GRU-32 model with the YCIFAR-10 dataset.
Figure~\ref{fig:rate_gain} shows the rate gain~\eqref{eq:rate_gain} versus accuracy for LDPC source coding rates $3/4$, $1/2$, and $1/4$, as well as when considering $4\times4$ and $2\times2$ DCT truncation, and with various number of bitplanes.  
The curves show that bitplane-based LDPC coding leads to reduced coding rates, while still maintaining the classification accuracy. For instance, an accuracy of $23\%$ can be achieved with a rate as low as $0.016$ bpp, and an accuracy of about $49\%$ is achieved with $50\%$ of data reduction. 



\subsection{Accuracy-complexity balance}
Table~\ref{tab:acc_weights} shows the number of learnable parameters and the achieved accuracy for various models on the MNIST dataset. The comparison includes the proposed DCT-LDPC scheme, the DCT with the Fully Connected model proposed in~\cite{fu2016using}, as well as Huffman with VGG and ResetNet models considered in~\cite{remy1}. We observe that our approach achieves a higher classification accuracy than existing methods while requiring approximately $1842$ fewer weights compared to ResetNet and $7187$ fewer weights compared to VGG-11. Even on the CIFAR-10 dataset with $24$ bitplanes (not shown in the Table), our proposed setup achieves an accuracy of $54\%$ while requiring only  $85k$ learnables. 
We conclude that the proposed method offers a much better balance between learning accuracy and complexity compared to state-of-the-art methods in~\cite{remy1} and~\cite{fu2016using}.

\begin{table}[htbp]
    \centering
     \caption{Reported Accuracy vs weights for MNIST dataset}  
    \begin{tabular}{lcc}
    \toprule

       \textbf{Setup} &  \textbf{Best accuracy $\uparrow$} & \textbf{No. of learnables $\downarrow$}\\
        \midrule
        Huffman on UVGG11\cite{remy1}& 0.8323 & $\sim 35$M\\
        Huffman on UResnet \cite{remy1}& 0.6313 & $\sim 5$M \\
      \cmidrule(lr){1-3}
        DCT on FullyConnected\cite{fu2016using} & 0.9000 & $1$M \\
     \cmidrule(lr){1-3}
        DCT-LDPC on 12 units GRU& \textbf{0.9060}  & \textbf{19k}\\
        DCT-LDPC on 32 units GRU& \textbf{0.9237}  & \textbf{52.6k}\\
        \bottomrule
        
    \end{tabular}

    \label{tab:acc_weights}
\end{table}

\subsection{Why LDPC codes work well?}
We now provide a summary of explanations about the improved accuracy achieved when learning on LDPC-coded data compared to other entropy-coding techniques. 

\subsubsection*{a) Sparsity}
LDPC codes are sparse in the sense that the matrix $H$ contains a small number of non-zero components. 
%
Therefore, each syndrome bits depends on a small subset of bits of the original image,   
which encapsulates the local features and patterns within the image. Therefore, the syndrome can be seen as a diverse set of characteristics that collectively represent the image.
\subsubsection*{b) Class-specific information preservation}
The LDPC coding operation can be regarded as a projection into a lower-dimensional syndrome space.  From the t-SNE analysis, it is apparent that images belonging to the same class frequently have similar pixel patterns. Linear projection realized by LDPC encoding preserves these similarities in the syndrome space. 

\subsubsection*{d) LDPC-GRU pair-specific combination}

The GRU model shows a suboptimal performance for classification over images without coding. However, it provides a very good classification performance when applied over LDPC codes. This may be attributed to the ability of GRU models to infer the structure of the LDPC coding matrix. The iterative (or recurrent) nature of the GRU model may closely mirror that of LDPC decoders such as belief propagation and min-sum algorithms.

\section{Conclusion}
\label{conc} 

In this paper, we conducted a rigorous analysis on the use of LDPC codes for entropy coding in the context of image classification without any prior decoding.  Our results demonstrate that combining LDPC codes with GRU models achieves impressive classification performance even with a lightweight GRU model.
We explored the influence of various code parameters on classification accuracy and showed that learning from a reduced number of bitplanes does not compromise accuracy. Additionally, the impact of the JPEG quality factor on classification outcomes proved to be significant. Our analysis further indicates that regular and irregular LDPC codes yield comparable accuracy, with regular codes exhibiting a slight advantage.
Future work will aim to extend this investigation to other learning tasks, such as image retrieval and segmentation, while also focusing on the development of LDPC codes specifically optimized for learning from compressed data.


\ifCLASSOPTIONcaptionsoff
  \newpage
\fi

\bibliographystyle{IEEEtran}
\bibliography{IEEEabrv,Bibliography}

\vfill
\end{document}